\documentclass{article}
\usepackage[utf8]{inputenc} \usepackage{siunitx} \usepackage{mathrsfs}
\usepackage{subfig} \usepackage{lscape} \usepackage{eucal}
\usepackage{afterpage,listings,multirow} \usepackage{relsize}
\usepackage{natbib} 
\usepackage{hyperref}
\usepackage{graphicx}
\usepackage{amsmath,bm,amssymb,mathtools} \usepackage[table]{xcolor}

\title{Joint models as latent Gaussian models - not
	reinventing the wheel}
\author{J.\ Van Niekerk, H.\ Bakka \& H.\ Rue\\
	CEMSE Division\\
	King Abdullah University of Science and Technology\\
	Saudi Arabia}

\date{January 2019}
\begin{document}
	\maketitle
	
	\begin{abstract}
		Joint models have received increasing attention during recent
		years with extensions into various directions; numerous hazard
		functions, different association structures, linear and non-linear
		longitudinal trajectories amongst others. Many of these resulted
		in new R packages and new formulations of the joint model.
		However, a joint model with a linear bivariate Gaussian
		association structure is still a latent Gaussian model (LGM) and
		thus can be implemented using most existing packages for LGM's. In
		this paper, we will show that these joint models can be
		implemented from a LGM viewpoint using the \textit{R-INLA}
		package. As a particular example, we will focus on the joint model
		with a non-linear longitudinal trajectory, recently developed and termed the
		partially linear joint model. Instead of the usual spline
		approach, we argue for using a Bayesian smoothing spline framework for the
		joint model that is stable with respect to knot selection and
		hence less cumbersome for practitioners.
	\end{abstract}

	\section{Introduction}
	
Latent Gaussian models (LGM's) is a group of models that contains most
statistical models used in practice. Indeed, most generalized linear mixed models (GLMM's) and general additive models (GAM's) that we are able to perform inference with, are  
examples of LGMs.  In the context of
joint models, this viewpoint has largely been under-presented and
merely mentioned in \cite{martino2011}. A joint model is unique in the
sense that there are two different likelihoods and shared random
effects in the model. Extensions of linear joint models like spatial
random effects, non-linear trajectories and multiple end-points
amongst others, are used in the context of joint models to address
certain practical challenges. Each of these new joint models is still
a latent Gaussian model and thus no special implementation package is
needed for each one. The \textit{R-INLA} package based on the INLA
methodology \citep{rue2009}, has been used extensively for latent
Gaussian models and could thus be used for joint models as well. Most
longitudinal likelihoods and hazard assumptions can be facilitated in
this framework, leaving no need for developing a new joint model for
each set of assumptions. As a particular example, we will focus on the
recently proposed partially linear joint model by 
\cite{kim2017}.\\ \\
Non-linear or partially linear joint models, in particular, is a
natural extension to the linear joint model since this is often the
case in real datasets. \cite{kim2017} introduced a
frequentist approach to fit a partially linear joint model using
splines and presented a selection method for the knot set based on
some model selection metrics. A Bayesian P-splines approach is adopted
in a joint model framework by \cite{kohler2017} where the number of
knots are also based on the value of some model selection metrics. The
approach proposed by \cite{kohler2017} uses the R package
\textit{bamlss} and the authors commented that the implementation of
this model is not computationally feasible. \\ \\ In this paper, we
present a Bayesian approach embedded within the \textit{R-INLA}
package \citep{rue2009} to fit a partially linear/non-linear joint
model, without the burden of choosing a specific set of knots. We use
a Bayesian smoothing spline model described by \cite{lindgren2008} and
\cite{yue2014}, that is the solution of a stochastic
differential equation (SDE) resulting in a second-order random walk
(see \cite{lindgren2008} and \cite{simpson2012} for further details)
in contrast to the semi-parametric Bayesian method proposed by
\cite{rizopoulos2011} that also depends on knot selection. Using this
methodology, the model is stable with regards to the choice of the
number and placement of the knots needed to form a spline basis as in
\cite{kim2017}, since it is a continuous time model. Additionally, our approach introduces a hyperparameter
pertaining to the spline component that is interpretable and can be
used to assess the appropriateness of the non-linear component. \\ \\
In Section \ref{jointmodelsection}, we present the partially linear
joint model as defined by \cite{kim2017}. Also, we present various
forms of the linear shared random effect that can be fitted using
\textit{R-INLA}, but not with most of the other available packages for
joint models. The Bayesian smoothing spline is discussed in
Section \ref{splines}. Latent Gaussian models and a synopsis of the
INLA methodology underpinning the \textit{R-INLA} package is presented
in Section \ref{lgmjointsec}. In this section, we also discuss how
joint models fit into the LGM framework. In Section
\ref{applicationsection}, we present an example of our approach and
compare it to that presented in \cite{kim2017} using the simluated PSA
dataset presented in the \textit{jplm} function in the \textit{JointModel} package. The paper is concluded
by concluding remarks in Section \ref{conclusionsection}.

\section{Partially linear joint model}\label{jointmodelsection}

A joint model consists of two marginal models, linked by shared,
correlated random effects. The motivation for the construction of such
a model is foundin the biological or physical process generating
the data, since multiple types of data generated by the same
individual is inherently correlated. The joint modeling of time to
event and longitudinal data is a fundamental tool in these type of
studies since insights about the survival component can be gained from
the longitudinal series (see
\cite{wulfsohn1997,hu2003,tsiatis2004,guo2004} amongst others for more
details). This is especially beneficial in studies where the events
are lengthy to observe or scarce. Usually, the model is constructed as
the combination of a longitudinal model to analyze data measured at
multiple time points based on the same investigative subject and a
survival model for the time to event data. This setup is quite common
in most studies where subjects are followed two-fold, biomarkers are
collected at multiple time points to investigate the behaviour of some
physical process (usually after some intervention/treatment) as well
as the absence/presence of a certain linked event (usually a relapse
or a fatal event).\\ \\ The models are jointly fitted by sharing a set
of random effects from the longitudinal submodel to the survival
submodel. This provides insights into the biological process acting as
the driving force behind various diseases such as prostate cancer
\citep{serrat2015}, ovarian cancer \citep{huang2018}, AIDS
\citep{guo2004,huang2018}, Dermatomyositis \citep{van2018} and Renal
disease \citep{rizopoulos2011}, to mention but a few. The exact form of
the shared random effect can vary. The most popular form currently
used is a linear random effect in time as the sum of a random
intercept and random slope over time, as implemented in the R packages
\textit{JMbayes, JointModel} and the function \textit{jplm} of which the latter can
incorporate a non-linear trajectory over time, in the longitudinal
submodel. Both Bayesian and frequentist methods have been developed
for joint models as summarized in the aforementioned packages, amongst
others. The linear shared random effect assumption has recently been
challenged by \citep{andrinopoulou2016}.
In this paper, however, we will focus on the case of linear shared
random effects.\\ \\
We denote $\pmb{y}$ and $\pmb{s}$ as the response vectors of the
longitudinal and survival submodels, respectively. Additionally,
$\pmb{X}$ and $\pmb{Z}$ is a set of available covariates for the
longitudinal and survival submodels, respectively.
\subsection{Longitudinal submodel}\label{longsec}
In various real-life situations, numerous datapoints are collected
from the same individual at different timepoints. This forms a
longitudinal series of data and cannot be modelled using standard
techniques like generalized linear models since the assumption of
independent and identically distributed observations does not hold.
Instead, conditional on the subject and/or group-specific random
effects, the observations are independent and identically distributed
in the context of a generalized linear mixed model. \\ \\ For each individual $i,i=1,...,N$ we have a vector of observations
$y_{ijk}=y_{ij}(t_{ijk})$ at various timepoints $t_{ijk}$, for groups $j=1,...,N_t$ where $ k=1,...,N_{ij}$, such that $\sum_i\sum_j N_{ij}=N_L$.
The longitudinal submodel
is a generalized mixed model for the longitudinal outcome in
continuous time. We assume that the conditional longitudinal outcomes
$y_{ijk}|(\pmb{\beta},\pmb{X}_{ijk},u_{ijk})$ are conditionally independent and follow a well-defined
distribution, $\mathscr{G}$ with some density function $g$, linear
predictor $\eta^L$ and hyperparameters $\pmb\theta_L$. In practice, a
Gaussian likelihood is often assumed, although this is not necessary
and any well-defined distribution can easily be facilitated in our
computational procedure. The longitudinal submodel is as follows:
\begin{eqnarray}
y_{ijk}|(\pmb{\beta},\pmb{X}_{ijk},u_{ijk})\sim
\mathscr{G}(\eta^L=\alpha(t_{ijk})+\pmb{\beta}^{T}\pmb{X}_{ijk}+u_{ijk})
\label{longsub}
\end{eqnarray}
In essence, the submodel is composed by a set of fixed effects,
$\pmb{\beta}^T\pmb{X}_{ijk}$, and a set of random effects,
$\alpha(t_{ijk})+u_{ijk}$. In this specification, $\pmb{\alpha}$
denotes the longitudinal trajectory which can assume any form, also
non-linear, with hyperparameters $\pmb{\theta}_{\pmb\alpha}$. The
random effects $\pmb{u}$ are the shared components, linear in time,
which forms the basis of the joint model. Specifically, we formulate,
\begin{equation}
u_{ijk}=u_{ij}(t_{ijk})=w_{ij}+v_{ij}t_{ijk}
\label{sharedrandomeff}
\end{equation} 
where $w_{ij}$ and $v_{ij}$ follow a bivariate Gaussian distribution
with zero mean, precision matrix $\pmb{Q}_u$ (inverse covariance matrix) and correlation
coefficient $\rho$. This Gaussian assumption is mostly used in literature.
\subsection{Survival submodel}\label{survsec}
Survival datasets are unique in the sense that the response for
subject $l$ consists of an event time, $s_l$, as well as a censoring
variable, $c_l$, to indicate if the time was censored $(c_l=0)$ or not
$(c_l=1)$. Right censoring is most commonly found in practice since
this results from terminating a study before all subjects experienced
the event. Let $s^*_l$ be the event time, if the subject experienced
the event $(c_l=0)$ and suppose $s^X$ is the last timepoint in the
study period, then, in the case of right censoring,
$s_l=\min(s^*_l,s^X)$. The construction of the time variable $s_l$, is
slightly different under different censoring schemes, so we will focus on right censoring in this paper.\\ \\The
specification of the form of the baseline hazard function $h_0(t)$ can
be achieved parameterically (exponential (constant hazard), Weibull
(monotonic hazard), log-Gaussian or log-Logistic (non-monotonic)
baseline hazard function) or semi-parametrically (Cox piecewise
constant model). Each of the afore-mentioned models can be used in our
computational procedure, so we will propose the survival submodel in a
general form and later present the specific details for each case. The
survival submodel is defined using the hazard rate as:
\begin{equation}
h_l(s)=h_0(s)\exp\left(\pmb{\gamma}^T\pmb{Z} +
\pmb{\nu}\circ(w_{l},v_{l}s)+m_l\right).
\label{survsub}
\end{equation}
and some hyperparameters $\pmb{\theta}_S$. If
$\pmb{\nu}\circ(w_{l},v_{l}s)$ is independent of time, then we have a
proportional hazards model. The plausibility of proportional hazards
should be investigated using exploratory analysis of the empirical
survival/hazard curves. The random effect $m_l$ is used to model
subject-specific variability in the survival time, often called a
frailty component, resulting in a frailty variable $\exp(m_l)$. The
association between the longitudinal and survival sub-models is
established by the term $\pmb{\nu}\circ(w_{l},v_{l}s)$.
\subsection{Possible linear association structures}
The joint model is solely developed based on the shared effects from
both submodels. A subset of the random effects in the longitudinal
model enter the hazard rate model through the well-defined combination
$\pmb{\nu}\circ(w_{l},v_{l}s)$ as in \eqref{survsub}. This combination
can be time-dependent, resulting in an accelerated failure time model.
The functional form of $\pmb{\nu}\circ(w_{l},v_{l}s)$, where $\circ$ is the component-wise product, can assume
various structures, some commonly found are summarized below
\citep{henderson2000}:
\begin{eqnarray}
\pmb{\nu}\circ(w_{l},v_{l}s)&=&\nu
w_l \label{share1}\\
\pmb{\nu}\circ(w_{l},v_{l}s)&=&\nu(v_ls)\label{share2}\\
\pmb{\nu}\circ(w_{l},v_{l}s)&=&\nu(w_l+v_ls)\label{share3}\\
\pmb{\nu}\circ(w_{l},v_{l}s)&=&\nu_1w_l+\nu_2(v_ls)\label{share4}
\end{eqnarray}
With the addition of \eqref{share3} in the \textit{R-INLA} package, all these functional forms, \eqref{share1}-\eqref{share4}, can be assumed in \textit{R-INLA}, while most
currently available R-packages for joint models, as well as
\textit{jplm}, can only facilitate \eqref{share3}. Within our
computational framework, the frailty variable should have a
log-Gaussian distribution, a priori. The log-gamma frailty has been
included in a test version and is discussed thoroughly in
\cite{martins2014}. In this paper, we will only focus on log-Gaussian
frailties to not distract from the main aim.
\section{Bayesian smoothing spline model}\label{splines}
As noted previously, the main aim of this paper is to formulate a
partially/non-linear joint model that can capture non-linear
trajectories presented by the data as an LGM. Traditionally, these
spline models have been based on selecting a number of knots or basis
functions, $B_k$, and then formulating a dependency structure through
coefficients $\lambda_k$, to give the random effect
$$\alpha(t) =  \pmb{B}(t) \pmb{\lambda}.$$
The choice of the placement and number of knots has been addressed in
various different ways. \cite{zhou2001} proposed a knot
relocation and search method instead of the stepwise addition and
deletion approach. A two-stage knot selection approach using wavelet
decomposition and then statistical model selection techniques was
proposed by \cite{he2001}, while
\cite{spiriti2013} introduced a stochastic search algorithm as an
improvement on multivariate adaptive regression splines (MARS)
\citep{friedman1991} to produce a near-optimal knot set in the squared
error sense. A Bayesian approach based on the joint posterior of the
placement and number of knots using piecewise polynomials is presented
in \cite{denison1998}, while
\cite{leitenstorfer2007} proposed using boosting techniques with
radial basis functions. Irrespective of the method used to find a knot
set, the main issue is that the number and locations of knots or basis
functions can change the model in fundamental ways. The reason the
model is unstable with respect to the choice of knots, is that the
covariance structure is built on the spline coefficients $\lambda$,
instead of on the spline $\alpha(t)$.\\ \\ A different approach to spline models is implemented in the
\textit{R-INLA} package and described by \cite{lindgren2008} and \cite{yue2014}. This approach
is based on finite element methods, frequently used in numerics and
mathematical modelling in general, where the focus is on approximating
some continuous spline $\alpha(t)$ on a discrete set of knots. The
covariance structure on $\lambda$ is derived by approximating the
desired covariance structure of $\alpha(t)$, following the theory of
numerical discrete approximations to continuous equations. Different
choices of knots or basis functions will approximate the same
continuous model, and, as the number of knots grow large, become
stable (and converge to the continuous model), contrary to most used
methods for spline regression.\\ \\ For the second order random walk, the continuous SPDE model is
$$\alpha''(t) = \mathcal W, $$ 
where $\alpha''$ denotes the second derivative and $\mathcal W$ the Gaussian white noise
process. 
For regular intervals, this can be approximated by
\begin{equation}
\alpha(t-1) -2 \alpha(t) + \alpha(t+1) \,{\buildrel d \over =}\ w_t, 
\label{rw2}
\end{equation}
where $w_t$ is Gaussian white noise with precision $\tau_\alpha$. The
use of irregular locations is found in \cite{lindgren2008}, and all
good approximations (the definition of ``good'' is studied in
numerical mathematics) will give very similar models. The next
modelling challenge with the random walk order 2, is that the size of
the spline (the range of values the spline can take) is difficult to
interpret, and it depends on the total number of observations. 
This
challenge was resolved by \cite{sorbye2014scaling},
and is implemented in R-INLA through the scale.model option.
With this approach, we can interpret the ``size'' of the
spline to be the overall deviation from a straight line (the straight
line has second derivative equal to zero).

\section{Latent Gaussian joint model}\label{lgmjointsec}
In this section we will briefly present the concept of latent Gaussian
models and the INLA methodology. Preference is given to the details
useful for this paper, further details can be found in \cite{rue2009}.
Joint models as presented in this paper, are shown to be LGM's and
hence fit into the INLA framework.

\subsection{Latent Gaussian models and INLA}\label{lgmsection}

Hierarchical Bayesian additive models are widely used in various
applications. A specific subset of Bayesian additive models is the
class of latent Gaussian models (LGM). An LGM can be efficiently
modelled using the INLA methodology implemented in the \textit{R-INLA}
package. This class comprises of well-known models such as mixed
models, temporal and spatial models. An LGM is defined as a model
having a specific hierarchical structure, as follows: The likelihood
is conditionally independent based on the likelihood parameters
(hyperparameters), $\pmb{\theta}$ and the linear predictors,
$\eta_i$, such that the complete likelihood can be expressed as
\begin{equation}
\pi(\pmb{y}|\pmb{\eta},\pmb{\theta})=\prod_{i=1}^{N}
\pi(y_i|\eta_i(\pmb{\mathcal{X}}),\pmb{\theta}).
\end{equation}
The linear predictor
is formulated as follows:
\begin{equation}
\eta_i=\beta_0+\pmb{\beta}^T\pmb{X}_i+\pmb{u}_i(\pmb{z}_i)+\epsilon_i
\label{additive predictor}
\end{equation}
where $\pmb{\beta}$ represent the linear fixed effects of the
covariates $X$, $\pmb{\epsilon}$ is the unstructured random effects
and $\pmb{\gamma}$ represents the known weights of the unknown
non-linear functions $\pmb{u}$ of the covariates $\pmb{z}$. The
unknown non-linear functions, also known as structured random effects,
$\pmb{u}$ include spatial effects, temporal effects, non-seperable
spatio-temporal effects, frailties, subject or group-specific
intercepts and slopes etc. This class of models include most models
used in practice since time series models, spline models and spatial
models, amongst others, are all included within this class. The main
assumption is that the data, $\pmb{Y}$ is conditionally independent
given the partially observed latent field, $\pmb{\mathcal{X}}$ and some
hyperparameters $\pmb{\theta}_1$. The latent field $\pmb{\mathcal{X}}$ is
formed from the structured predictor as
$(\pmb{\beta},\pmb{u},\pmb{\eta})$ which forms a Gaussian Markov
random field with sparse precision matrix $\pmb{Q}(\pmb{\theta}_2)$,
i.e.\ $\pmb{\mathcal{X}}\sim N(\pmb{0},\pmb{Q}^{-1}(\pmb{\theta}_2))$. A prior,
$\pmb{\pi}(\pmb{\theta})$ can then be formulated for the set of
hyperparameters $\pmb{\theta}=(\pmb{\theta}_1,\pmb{\theta}_2)$. The
joint posterior distribution is then given by:
\begin{equation}
\pmb{\pi}(\pmb{\mathcal{X}},\pmb{\theta})\varpropto\pmb{\pi}(\pmb{\theta})\pmb{\pi}
(\pmb{\mathcal{X}}|\pmb{\theta})\prod_{i}\pi(Y_i|\pmb{\mathcal{X}},\pmb{\theta})
\label{postINLA}
\end{equation}
The goal is to approximate the joint posterior density \eqref{postINLA} and subsequently compute the marginal posterior densities,
$\pmb{\pi}(\mathcal{X}_i|\pmb{Y}),i=1...n$ and
$\pmb{\pi}(\pmb{\theta}|\pmb{Y})$. Due to the possibility of a non-Gaussian likelihood, the Laplace approximation to approximate this analytically intractable joint posterior density. The sparseness assumption on the precision of
the latent Gaussian field ensures efficient computation \citep{rue2005} .

\subsection{Joint models as latent Gaussian models}

In this section, we will briefly show that the joint model is indeed
an LGM as defined in Section \ref{lgmsection}. The likelihood for the
survival submodel from Section \ref{survsec} is
\begin{equation*}
\pi_S(\pmb s|\pmb{Z},\pmb{\gamma})=\prod_{l=1}^{N}\pi_l(s|\pmb{Z},\pmb{\gamma})
=\prod_{l=1}^N f_l(s)^{c}[1-F_l(s)]^{1-c}
\label{longlik}
\end{equation*}
where $f_l(s)=h_l(s)\exp\left(-\int_0^s h_l(u)du\right)$ from
\eqref{survsub}. The likelihood for the longitudinal biomarker from
Section \ref{longsec} is
\begin{equation*}
\pi_L(\pmb{y}|\pmb{X},\pmb{\beta})=\prod_{i=1}^{N_L} g(y_i).
\label{survlik}
\end{equation*}
The associated linear predictors are
\begin{eqnarray}
\eta^S&=&\pmb\gamma^T\pmb Z+\pmb{\nu}\circ(w_{l},v_{l}s)+m_l\notag\\
\eta^L&=&\alpha(t)+\pmb\beta^T\pmb X+u.\label{jointexample}
\end{eqnarray}
Note that each longitudinal observation is connected to the latent
field through the linear predictor $\eta^L$ in \eqref{longsub} and each
survival time through the linear predictor $\eta^S$ in \eqref{survsub}.
Now consider the hyperparameters
$\pmb\theta=\{\pmb\theta_l,\pmb\theta_S,\pmb\theta_m,\tau_\alpha,\pmb\nu,\pmb
Q_u,\rho\}$, the latent field
$\pmb{\mathcal{X}}=(\pmb\eta^L,\pmb\eta^S,\pmb\beta,\pmb\gamma,\pmb{v},\pmb{w},\pmb{m},\pmb{\alpha})$
conditioned on $\pmb\theta$ has a Gaussian distribution with precision
matrix $\pmb Q(\pmb\theta )$. From this construction of the latent
field, the observations $(\{y_{ijk}\},\{s_l,c_l\})$ have a complete
likelihood that depends on $\pmb{\mathcal{X}}$ only through one of the
linear predictors, $(\{\pmb\eta^L\},\{\pmb\eta^S\})$. Finally the
hyperparameters $\pmb\theta$ are assigned a prior distribution
$\pi(\pmb\theta)$. Hence, the partially linear joint model as
presented here, is an LGM and we can thus use the INLA methodology for
efficient Bayesian inference. A simple example using a simulated dataset with a non-linear longitudinal trend is available in Appendix \ref{appendixjoint} for illustration purposes.

\section{Example: PSA study}\label{applicationsection}
In prostate cancer studies, Prostate-specific Antigen (PSA) has been
identified as a biomarker for the status of prostate cancer. High
levels of PSA are indicative of increased risk of prostate cancer or
recurrence. Radiation therapy is a common course of treatment often
prescribed for patients with prostate cancer. If successful, the PSA
levels are expected to drop and remain at a low level. On the
contrary, PSA levels will drop initially and then rise again
\citep{zagars1995}. Hence, it is desirable to develop a flexible model
to capture this nonlinear temporal trend of PSA levels per patient. A
challenge is that the follow-up of PSA is stopped when salvage hormone
therapy is initiated, which is known to change the PSA level or when
prostate cancer recurred, resulting in possibly informative drop-out.
If this informative drop-out is unaccounted for, it can lead to
considerable bias in the PSA trajectory estimation. The objective of
this analysis is thus, to identify the trajectory of post-radiation
PSA change, while correctly accounting for the informative drop-out.
More details about the clinical impact of such a study can be found in
\cite{proust2009}.

\subsection{Partially linear joint model}

In \cite{kim2017} a partially linear joint model is proposed utilizing
a spline component to capture the non-linear trajectory. They
developed a procedure using BIC for the knot selection needed to fit
this spline. In this paper, however, we use the Bayesian 
smoothing spline model as presented in Section \ref{splines} to
capture the non/semi-linear trajectory of PSA levels using INLA. This
approach facilitates a computationally efficient and user-friendly
implementation of these types of models, and is stable with regards to
the knot set. This approach produces reproducible and reliable
results. The joint model under consideration in this application from
\eqref{jointexample}, is:
\begin{eqnarray}
\log(\text{PSA})(t)&=&\eta^L+\epsilon(t)\notag \\
h(s)&=&h_0(s)\exp(\eta^S)\notag
\end{eqnarray} 
where $\epsilon\sim N(0,\sigma^2_\epsilon)$. We assume a Weibull
baseline hazard function, hence $h_0(s)=\kappa s^{\kappa -1}$ which is
non-constant over time. The exponential baseline hazard function with
constant hazard can be achieved as a special case when $\kappa=1$. In
\cite{kim2017} the functional form $\pmb{\nu}\circ(w,vs)=\nu(w+vs)$ as
in \eqref{surv1} is used, which is the most commonly used form of
shared effects in joint models. This form has now been included in the
\textit{R-INLA} package using the model "intslope". To facilitate a more
general structure, we also consider
$\pmb{\nu}\circ(w,vs)=\nu_1w+\nu_2(vs)$ as in \eqref{surv2}, hence the
linear predictors are formulated as:
\begin{eqnarray}
\eta^L&=&\alpha(t)+\beta \log(\text{PSA}_\text{base})+w+vt\notag\\
\eta^S_{1}(s)&=&\gamma \log(\text{PSA}_\text{base})+\nu(w+vs)\label{surv1}\end{eqnarray}
and 
\begin{eqnarray}
\eta^L&=&\alpha(t)+\beta log(\text{PSA}_\text{base})+w+vt\notag\\
\eta^S_{2}(s)&=&\gamma log(\text{PSA}_\text{base})+\nu_1w+\nu_2(vs)\label{surv2}
\end{eqnarray}
where
$$\begin{bmatrix} w \\ v \end{bmatrix}\sim
N \begin{pmatrix} \begin{bmatrix}0 \\ 0 \end{bmatrix}, \begin{bmatrix}
\sigma^2_w & \rho\sigma_w\sigma_v\\ \rho\sigma_w\sigma_v &
\sigma^2_v \end{bmatrix} \end{pmatrix}$$ and $\alpha(t)$ is a
second order random walk model as described in Section \ref{splines}.
Within the INLA framework, the number of groups for the local spline
should be specified. This number should have minimal influence on the
estimated result due to the construction presented in Section
\ref{splines}. On the contrary, it is well-known that the number of
knots greatly influence the estimated spline using more traditional
methods as in \cite{kim2017}. This conjecture is further discussed in
teh presentation of the results for the dataset under discussion.

\subsection{Bayesian inference}

The linear predictors under consideration \eqref{jointexample}
contains various components of the latent field and also some
hyperparameters. The prior for the latent field is assumed to be
multivariate Gaussian. The regression coefficients for the fixed
affects are assigned vague independent Gaussian priors. Following
\cite{simpson2017}, we assign penalized complexity priors for the
hyperparameters in the model as far as possible.

The random walk order two model $\alpha(t)$ in \eqref{rw2} has one
hyperparameter, $\tau$ which is assigned a penalized complexity prior
with prior density
$$\pi(\tau_\alpha)=\lambda_\alpha\tau_\alpha^{-\frac{3}{2}}
\exp(-\lambda_\alpha\tau^{-\frac{1}{2}})$$ such that
$P(\frac{1}{\sqrt{\tau_\alpha}}>1)=0.01$, i.e.\
$\lambda_{\alpha}=ln(0.01)$, which is the Gumbel type 2 distribution.
The bivariate random effect $\begin{bmatrix} w \\ v \end{bmatrix}$
assumes a bivariate Gaussian prior with covariance matrix
$\tau^{-1}_{w,v}\pmb{R}^{-1},\pmb{R}\geq\pmb{0}$ with a penalized
complexity prior for $\tau_{w,v}$ as the Gumbel type 2 distribution
with parameter $\log(0.01)$, as well. \\ \\The motivation for employing
penalized complexity priors for the precision hyperparameters are
founded in the fact that the usual priors for the variance components,
i.e. independent inverse-gamma priors as in \cite{huang2018}, overfits
and cannot contract to the simpler model in which the respective model
component has trivial variance. This is especially important in the
case of joint models since the effect of overfitting is exacerbated by
the influence of the shared random effect on all the linear
predictors.

\subsection{Results}\label{secpsaresults}

\begin{figure}[h]
	\centering \includegraphics[width=16cm]{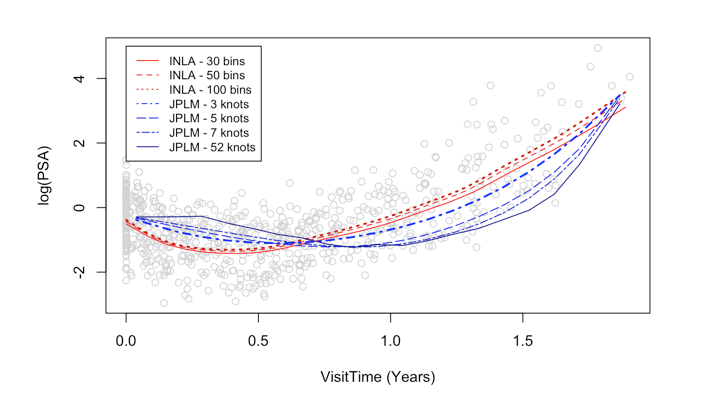}
	\caption{Estimated non-linear post PSA trajectory using
		\textit{R-INLA} and \textit{jplm}}
	\label{fignonlin}
\end{figure}

The two aforementioned models \eqref{surv1} and \eqref{surv2} were
both fitted using \textit{R-INLA} (for more details see the Appendix)
and \eqref{surv1} was also fitted using \textit{jplm} for comparison
purposes (the code is available in Web Appendix A). Firstly, the estimated post-treatment PSA trajectories are
presented in Figure \ref{fignonlin}. It is apparent that the number of
knots changes the shape of the estimated trajectory to a large extent.
For a low number of knots, the estimated trajectory is strictly convex
but as the number of knots increase, the trajectory contains concave
and convex parts. This challenge is not present in the trajectories
estimated using \textit{R-INLA}. Even for differing number of groups and
knot placement, the shape of the estimated trajectory is preserved. It
is clear from Figure \ref{fignonlin} that the trajectories estimated
from \textit{R-INLA} are supported by the data to a larger extent than
some of the trajectories estimated from \textit{jplm}. This behaviour
of a spline is inherent in the formulation and construction of the
spline model as a combination of basis functions with associated
random weights, as opposed to the formulation as presented in Section
\ref{splines} and implemented in \textit{R-INLA}.
Secondly, the resulting estimated joint model is presented. The
results are summarized in Table \ref{tableres1}. It is evident that
the two estimation procedures provide similar results although the
uncertainty from using \textit{R-INLA} is higher. This is an expected
result from a Bayesian viewpoint.
\begin{table}[h]
	
	\begin{tabular}{|c||c|c||c|c|}
		\hline
		\multirow{2}{6em}{\textbf{Parameter}} & \textbf{Posterior Mode} & \textbf{Posterior SD} & \textbf{Point estimate} & \textbf{Standard error} \\ \cline{2-5}
		& \multicolumn{2}{c||}{\textbf{Joint model 1 - INLA}} & \multicolumn{2}{c|}{\textbf{Joint model 1 - jplm}} \\ \hline
		$\beta$ & $0.450$ & $0.061$& \cellcolor{gray!10} $0.443$ & \cellcolor{gray!10}$0.004$ \\
		$\gamma$ & $0.743$ & $0.198$ & \cellcolor{gray!10}$0.742$ &\cellcolor{gray!10} $0.041$\\
		$\sigma^2_\epsilon$ & $0.091$ & $0.005$ & \cellcolor{gray!10}$0.089$ &\cellcolor{gray!10} $0.003$\\
		$\sigma^2_\alpha$ & $0.226$ & $0.143$ &\cellcolor{gray!10} NA &\cellcolor{gray!10} NA\\
		$\sigma^2_w$ & $0.342$ & $0.053$ & \cellcolor{gray!10}$0.328$ &\cellcolor{gray!10} $0.003$\\ 
		$\sigma^2_v$ & $0.216$ & $0.054$& \cellcolor{gray!10}$0.181$ &\cellcolor{gray!10} $0.002$  \\
		$\rho$ & $-0.131$ &$0.149$&\cellcolor{gray!10} $-0.172$ & \cellcolor{gray!10}$0.025$ \\ 
		$\nu$ & $0.921$ & $0.137$ &\cellcolor{gray!10} $1.122$ &\cellcolor{gray!10} $0.121$ \\
		$\kappa$ & $0.806$ & $0.092$ &\cellcolor{gray!10} NA &\cellcolor{gray!10} NA \\\hline  
	\end{tabular}
	\caption{Results for the PSA dataset using \textit{R-INLA} and
		\textit{jplm} for the specification in \eqref{surv1}}
	\label{tableres1}
\end{table}
It is quite clear from Table \ref{tableres1} that the hazard of
informative dropout is correlated with the longitudinal PSA biomarker
since $\nu=0.919$ with $95\%$ credible interval $(0.645;1.193)$. This
result confirms that the joint model approach is supported by the data
and should be preferred to the separate models. The structure of the
association term as in \eqref{surv1} is quite restrictive but has been
used extensively. We also investigate the possibility of changing the
association structure to \eqref{surv2} and present the results in
Table \ref{tableres2}.

\begin{table}[h]
	\begin{tabular}{|c||c|c|}
		\hline
		\multirow{2}{6em}{\textbf{Parameter}} & \textbf{Posterior Mode} & \textbf{Posterior SD} \\ \cline{2-3}
		& \multicolumn{2}{c|}{\textbf{Joint model 2 - INLA}} \\ \hline
		$\beta$ & $0.389$ & $0.098$ \\
		$\gamma$ & $0.698$ & $0.193$ \\
		$\sigma^2_\epsilon$ & $0.125$ & $0.007$ \\
		$\sigma^2_\alpha$ & $0.166$ & $0.105$ \\
		$\sigma^2_w$ & $0.201$ & $0.032$ \\
		$\sigma^2_v$ & $0.365$ & $0.203$ \\
		$\rho$ & $-0.431$ & $0.257$ \\
		$\nu_1$ & $1.025$ & $0.270$ \\
		$\nu_2$ & $0.562$ & $0.308$  \\ 
		$\kappa$ & $0.817$ & $0.093$ \\ \hline
		
	\end{tabular}
	\caption{Results for the PSA dataset using \textit{R-INLA} for the
		specification in \eqref{surv2}}
	\label{tableres2}
\end{table}

\begin{figure}[h]\vspace{-1.5cm}
	\includegraphics[width=7cm]{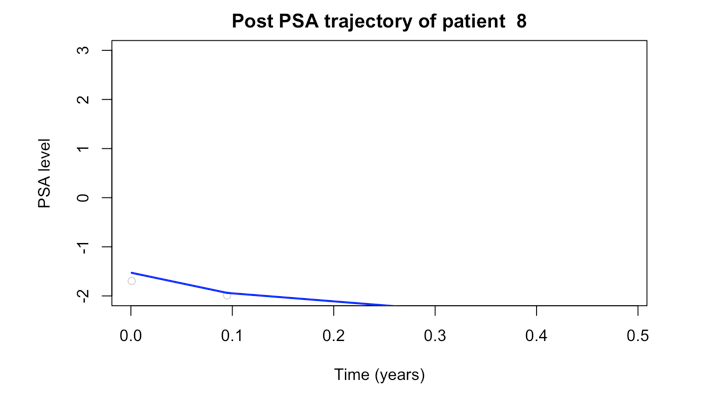}
	\includegraphics[width=7cm]{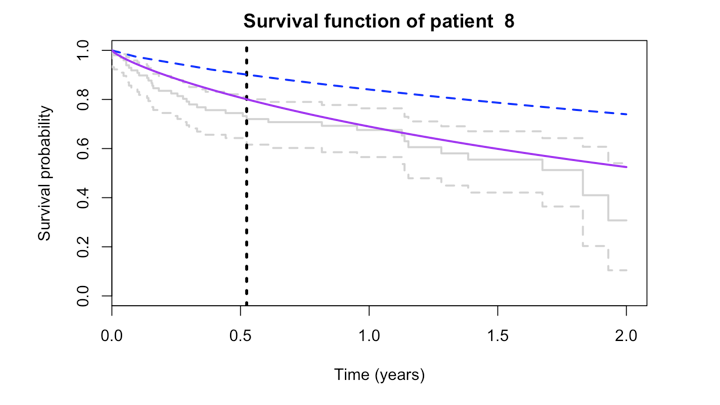}\\
	\includegraphics[width=7cm]{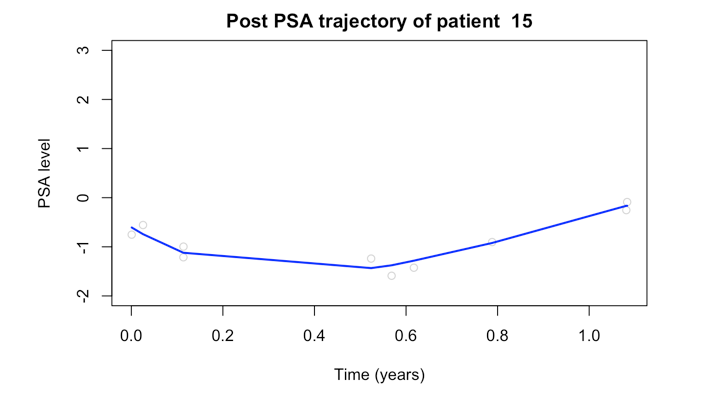}
	\includegraphics[width=7cm]{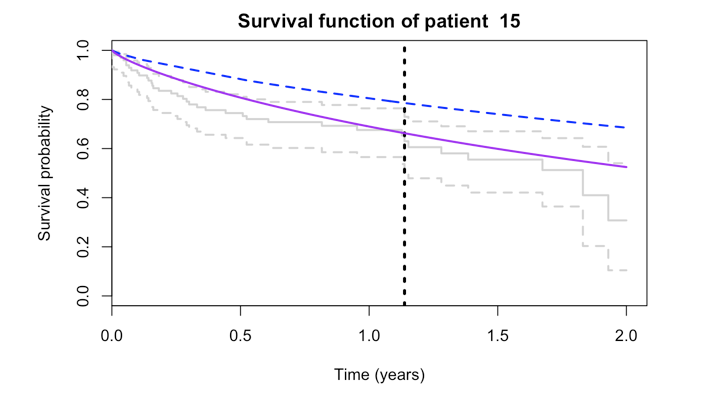}\\
	\includegraphics[width=7cm]{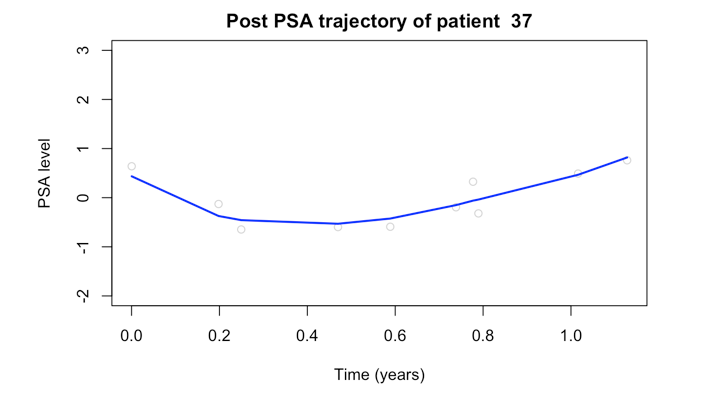}
	\includegraphics[width=7cm]{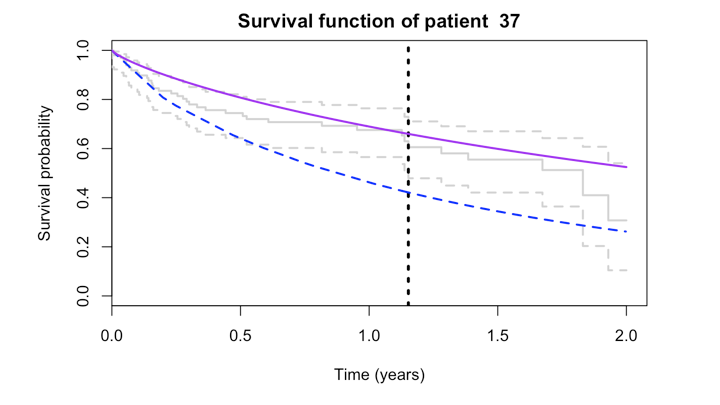}\\
	\includegraphics[width=7cm]{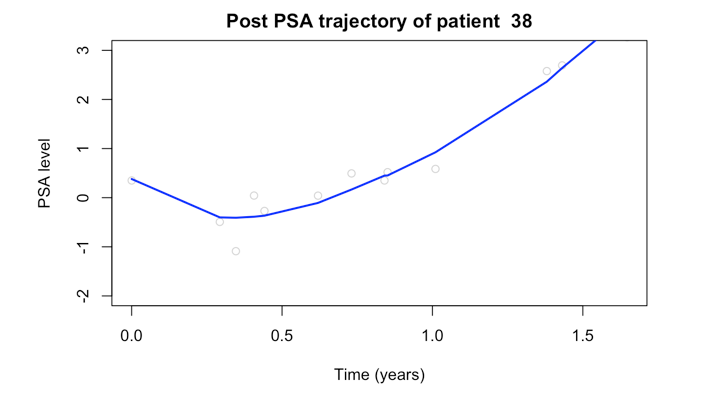}
	\includegraphics[width=7cm]{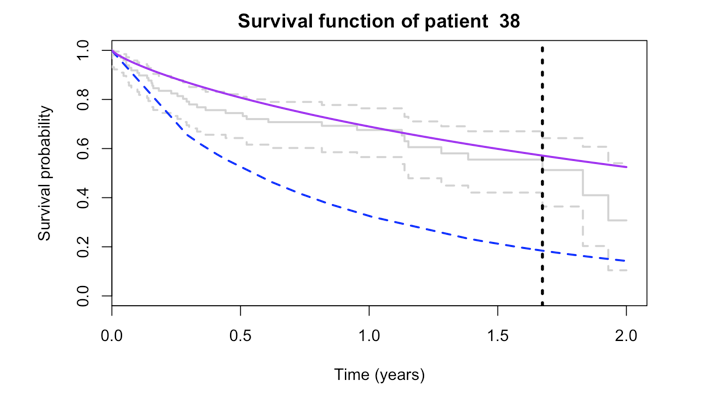}\\
	\includegraphics[width=7cm]{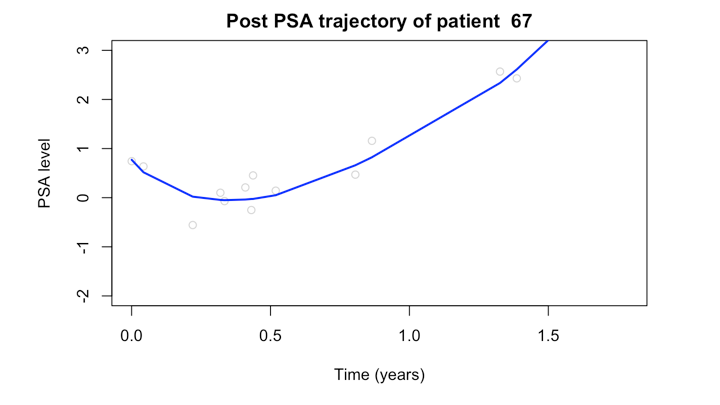}
	\includegraphics[width=7cm]{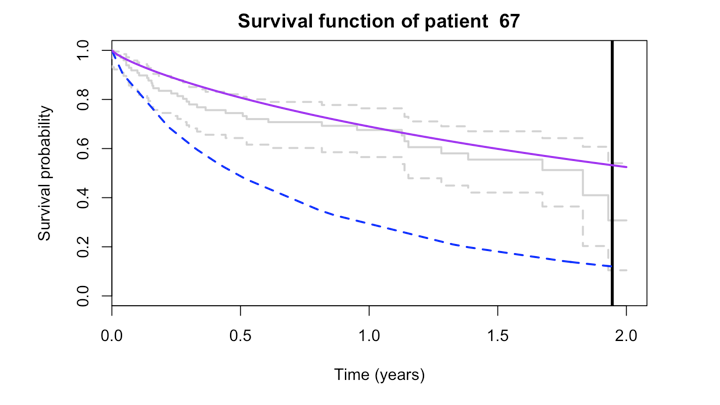}\\
	\includegraphics[width=7cm]{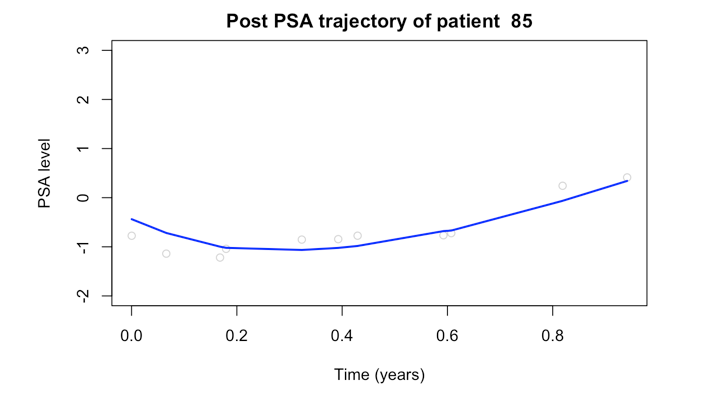}
	\includegraphics[width=7cm]{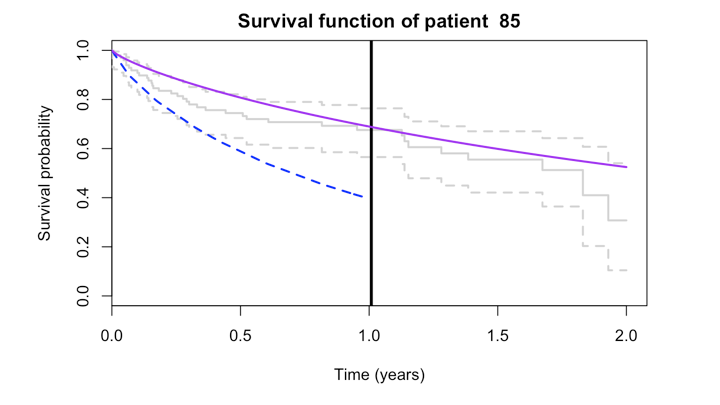}
	\caption{Post PSA trajectories and Survival functions for specific
		patients}
	\label{figpat}
\end{figure}

In comparison, the results between models 1 and 2 are very similar for
the fixed effects and variance hyperparameters. The interesting
difference between the two models as presented in Tables
\ref{tableres1} and \ref{tableres2}, is that the values of $\nu_1$ and
$\nu_2$ are quite different from each other, and from $\nu$ in Table
\ref{tableres1}. This implies that the structure of the shared effect
presented in \eqref{surv1} is not supported by the data in this
example and the more flexible model as in \eqref{surv2} should rather
be used. The model in \eqref{surv2} is not available in most of the
packages mentioned throughout the paper, but is feasibly implemented
in the \textit{R-INLA} package.\\ \\ 
In Figure \ref{figpat}, we present some of the longitudinal
trajectories and survival curves (or in the context of this
application, the non-dropout probabilities) for individual patients
based on \eqref{surv2}. The vertical line indicates the time at which
the dropout (solid) or censoring (dashed) occurred. The stepwise curve
is the Kaplan-Meier estimate of the survival curve for all patients,
the solid curve indicates the estimated mean survival curve from our
model and the dashed curve is the patient-specific survival curve.

\noindent The association between the PSA biomarker and the risk of
dropout is evident from Figure \ref{figpat}. Patients with the
distinctive decrease-increase behaviour are at higher risk of dropout
(lower survival function) and most eventually dropped out as indicated
by the solid vertical line. We have included two patients (8 and 15)
whose dropout times are censored but based on their PSA biomarker
levels, their survival functions are higher that the mean survival and
they would thus be considered for non-dropout. On the contrary,
patients 37 and 38 display the typical decrease-increase behaviour and
their survival functions indicate the higher probability of dropout.
The patient specific results can be used for dynamic predictions to
identify those patients who are most at risk of dropout, amongst other
things. The appropriateness of our proposed model is clear from the
detailed discussion of this specific example. The method can be
applied to various other datasets usually used in joint model analysis
using the \textit{R-INLA} package.

\section{Conclusion}\label{conclusionsection}

Joint models is one of the most common approaches used to analyze clinical time to event data. Consequently, various extensions and generalizations have been
developed, each with its own implementation structure. There are
various R packages available as mentioned, from both frequentist and
Bayesian viewpoints. In this paper, however, we showed that any joint
model with linear association structure, is indeed a simple latent
Gaussian model and all tools for LGM's can thus be applied in the
context of joint models. One of the most established and popular tools
for LGM's, is the INLA framework embedded in the \textit{R-INLA}
package. This affords the use of complicated joint models with relative ease, even as the models evolve in complexity. Model based evaluation of the assumptions, like the assumed association structure or non-linearity, is done with little effort within the \textit{R-INLA} framework since a multitude of joint model structures can be facilitated in this framework. \\ \\
As an example, we focused on a partially linear joint model with a
spline component to accommodate for non-linear longitudinal
trajectories. Instead of the usual splines approach with a set of
basis functions and corresponding regression coefficients. From a
Bayesian perspective, priors are usually assumed for the regression
coefficients. We proposed an alternative approach, that assumes priors
for the spline itself. This results in a spline component where
the user is relieved of the burden of knot selection. Subsequently, we
assumed penalized complexity priors to achieve shrinkage in the joint
model. The applicability of this proposal was illustrated using data
from a Prostate cancer study using PSA levels and time to dropout.\\ \\
Ultimately, the developments presented in this paper grants the application of complexities in joint models, such as non-linear or spatial components, not readily available for practical use in most other R packages. The proposed approach is useful and wieldy for practitioners and
statisticians alike, using the \textit{R-INLA} package for efficient
implementation.

\section*{Supplementary Materials}

Web Appendix A, referenced in Section~\ref{secpsaresults}, is available with
this paper at the Biometrics website on Wiley Online
Library.\vspace*{-8pt}

\bibliographystyle{apalike}
\bibliography{BioJ}


\section{Appendix}
\subsection{Computational considerations for joint models in using INLA}\label{appendixjoint}
	The likelihood of a joint model basically consists of two types of likelihoods and this can be facilitated in the INLA framework. It is essential to construct the response matrix and the covariate matrices correctly for the estimation procedure. For the purpose of this paper, we will present only the case where the joint model consists of longitudinal and survival submodels. This can be extended to include more marginal submodels in the case of multiple endpoint modeling. \\ \\
	Within the context of this paper, consider the following structured predictors of the longitudinal and survival submodels, respectively:
	\begin{eqnarray}
	\eta^L_{ijk}&=&\alpha(t_{ijk})+\pmb{\beta}^{T}\pmb{X}_{ijk}+w_{ij}+v_{ij}t_{ijk}\notag\\
	\eta^S_{l}(s)&=&\pmb{\gamma}^{T}\pmb{Z}_{l}+\pmb{\nu}\circ(w_{l},v_{l}s)\label{jointexample1}
	\end{eqnarray}
	Consider the case where the data consists of $N_i,i=1,...,N$ observations for each of the $N$ individuals, so that in total there are $N_L$ longitudinal observations and correspondingly $N_S=N$ event times and censoring indicators $(s_i,c_i),i=1,...,N$. The data is then composed as a list in which each variable consists of $N_L+N_S$ elements. To achieve this, we include zeros for fixed effects if the covariate is not included in that specific submodel and NA's for the random effects. In the case of \eqref{jointexample1}, the new response is defined as a list of the $y_{ijk}$ and $(s_i, c_i)$. The fixed effect covariates are constructed as $(\pmb{X},\pmb{0}_{1,...,N_S})$ and $(\pmb{0}_{1,...,N_L},\pmb{Z})$ while the random effects are constructed as $(\pmb{\alpha},\pmb{NA}_{1,...,N_S})$. \\ \\
	The main contribution in this area is the estimation of $\pmb{\alpha}$. Most of the commonly used  approaches to estimate the non-linear trend invloves the use of knots. This method was also used in \cite{kim2017}. In this paper we propose the use of a time-continuous spline model manifested as a second-order random walk presented in Section \ref{splines}.
\subsection{Example: Simulated joint model}\label{appsim}
	In this example we simulated data from the following scenario:
	\begin{eqnarray*}
	\eta^L(t)=t^2+v_i\notag\\
	\eta^S=\beta_Sv_i\notag
	\end{eqnarray*}where $v_i\sim N(0,\sigma^2_v)$ are the subject-specific random effects that are shared in this joint model. The aim of this example is to illustrate the practical method to fit a joint model in \textit{R-INLA}. The R code is available at \url{http://www.r-inla.org/examples/case-studies/van-niekerk-bakka-and-rue-2019}. 
\subsection{Example: PSA study - computational framework information}\label{apppsa}
	The R code used to obtain the results as presented in Section \ref{applicationsection} is available at \url{http://www.r-inla.org/examples/case-studies/van-niekerk-bakka-and-rue-2019}.\\ 
	The computational time needed was $83.2$ and $15.7$ seconds, respectively, for models 2 and 1 fitted using the \textit{INLA} package.
	The computer used is an Apple Macbook Pro i5 3.1GHz with 16GB 2133 MHz LPDDR3.

\end{document}